\documentclass[aps,prl,reprint,superscriptaddress]{revtex4-2}
\usepackage{blindtext}
\usepackage{graphicx}
\usepackage{amsmath}
\usepackage{amssymb}
\usepackage[english]{babel}
\usepackage{xcolor}
\usepackage{siunitx}
\usepackage[colorlinks=true, pdfstartview=FitV, linkcolor=blue,
citecolor=blue, urlcolor=blue]{hyperref}
\usepackage[capitalise]{cleveref}

\usepackage{float}

\makeatletter
\let\saved@includegraphics\includegraphics
\AtBeginDocument{\let\includegraphics\saved@includegraphics}
\renewenvironment*{figure}{\@float{figure}}{\end@float}

\makeatother

\makeatletter
\newcommand\footnoteref[1]{\protected@xdef\@thefnmark{\ref{#1}}\@footnotemark}
\makeatother

\begin{document}

\title{Phase driving  hole spin qubits}

\author{Stefano Bosco}
\email{stefano.bosco@unibas.ch}
\affiliation{Department of Physics, University of Basel, Klingelbergstrasse 82, 4056 Basel, Switzerland}
\author{Simon Geyer}
\affiliation{Department of Physics, University of Basel, Klingelbergstrasse 82, 4056 Basel, Switzerland}
\author{Leon C. Camenzind}
\altaffiliation[Present address: ]
{ RIKEN, Center for Emergent Matter Science (CEMS), Wako-shi, Saitama 351-0198, Japan}
\affiliation{Department of Physics, University of Basel, Klingelbergstrasse 82, 4056 Basel, Switzerland}
\author{Rafael S. Eggli}
\affiliation{Department of Physics, University of Basel, Klingelbergstrasse 82, 4056 Basel, Switzerland}
\author{Andreas Fuhrer}
\affiliation{IBM Research Europe-Zurich, Säumerstrasse 4, CH-8803 Rüschlikon, Switzerland}
\author{Richard J. Warburton}
\affiliation{Department of Physics, University of Basel, Klingelbergstrasse 82, 4056 Basel, Switzerland}
 \author{ Dominik M. Zumb\"uhl}
\affiliation{Department of Physics, University of Basel, Klingelbergstrasse 82, 4056 Basel, Switzerland}
\author{J. Carlos Egues}
\affiliation{Department of Physics, University of Basel, Klingelbergstrasse 82, 4056 Basel, Switzerland}
\affiliation{Instituto de Física de São Carlos, Universidade de São Paulo, 13560-970 São Carlos, São Paulo, Brazil}
\author{Andreas V. Kuhlmann}
\affiliation{Department of Physics, University of Basel, Klingelbergstrasse 82, 4056 Basel, Switzerland}
\author{Daniel Loss}
\affiliation{Department of Physics, University of Basel, Klingelbergstrasse 82, 4056 Basel, Switzerland}

\begin{abstract}
The spin-orbit interaction in spin qubits enables spin-flip transitions, resulting in Rabi oscillations when an external microwave field is resonant with the qubit frequency.
Here, we introduce an alternative driving mechanism of hole spin qubits, where a far-detuned oscillating field couples to the qubit phase. 
Phase driving at radio frequencies, orders of magnitude slower than the microwave qubit frequency, induces highly non-trivial spin dynamics, violating the Rabi resonance condition. By using a qubit integrated in a silicon fin field-effect transistor (Si FinFET), we demonstrate a controllable suppression of resonant Rabi oscillations, and their revivals at tunable sidebands. These sidebands enable alternative qubit control schemes using global fields and local far-detuned pulses, facilitating the design of  dense large-scale qubit architectures with local qubit addressability.
Phase driving also decouples Rabi oscillations from noise, an effect due to a gapped Floquet spectrum and can enable Floquet engineering high-fidelity gates in future quantum processors.
\end{abstract}

\maketitle

\paragraph{Introduction.|}

Spin qubits in hole quantum dots are emerging as top candidates to build large-scale quantum processors~\cite{doi:10.1146/annurev-conmatphys-030212-184248,scappucci2020germanium,fang2022recent,hendrickx2020four}.
A key advantage of hole spins is their large and tunable spin-orbit interaction (SOI) enabling ultrafast all-electrical qubit operations~\cite{Wang2022,Froning2021,watzinger2018germanium,Hendrickxsingleholespinqubit2019,hendrickx2020fast,maurand2016cmos,camenzind2021spin}, on-demand coupling to microwave photons~\cite{yu2022strong,PhysRevLett.129.066801,michal2022tunable}, even without bulky micromagnets~\cite{Philips2022,doi:10.1126/sciadv.abn5130,Noiri2022}. 
The large SOI of holes leads to interesting physical phenomena, such as electrically tunable Zeeman~\cite{PhysRevResearch.3.013081,PhysRevB.104.235303,hudson2022observation,PhysRevB.106.235408,PhysRevB.104.115425,Froning2021} and hyperfine interactions~\cite{PhysRevLett.127.190501,PhysRevB.78.155329,Prechtel2016}, or exchange anisotropies at finite~\cite{geyer2022two} and zero magnetic fields~\cite{Katsaros2020,PhysRevLett.129.116805}. These effects can be leveraged for quantum information processing, e.g., to define operational sweet spots against noise~\cite{PRXQuantum.2.010348,PhysRevApplied.18.044038,PhysRevB.105.075308,Piot2022,Wang2021,wang2022modelling}; to date their potential remains  largely unexplored.  

\begin{figure}[t!]
\centering
\includegraphics[width=\columnwidth]{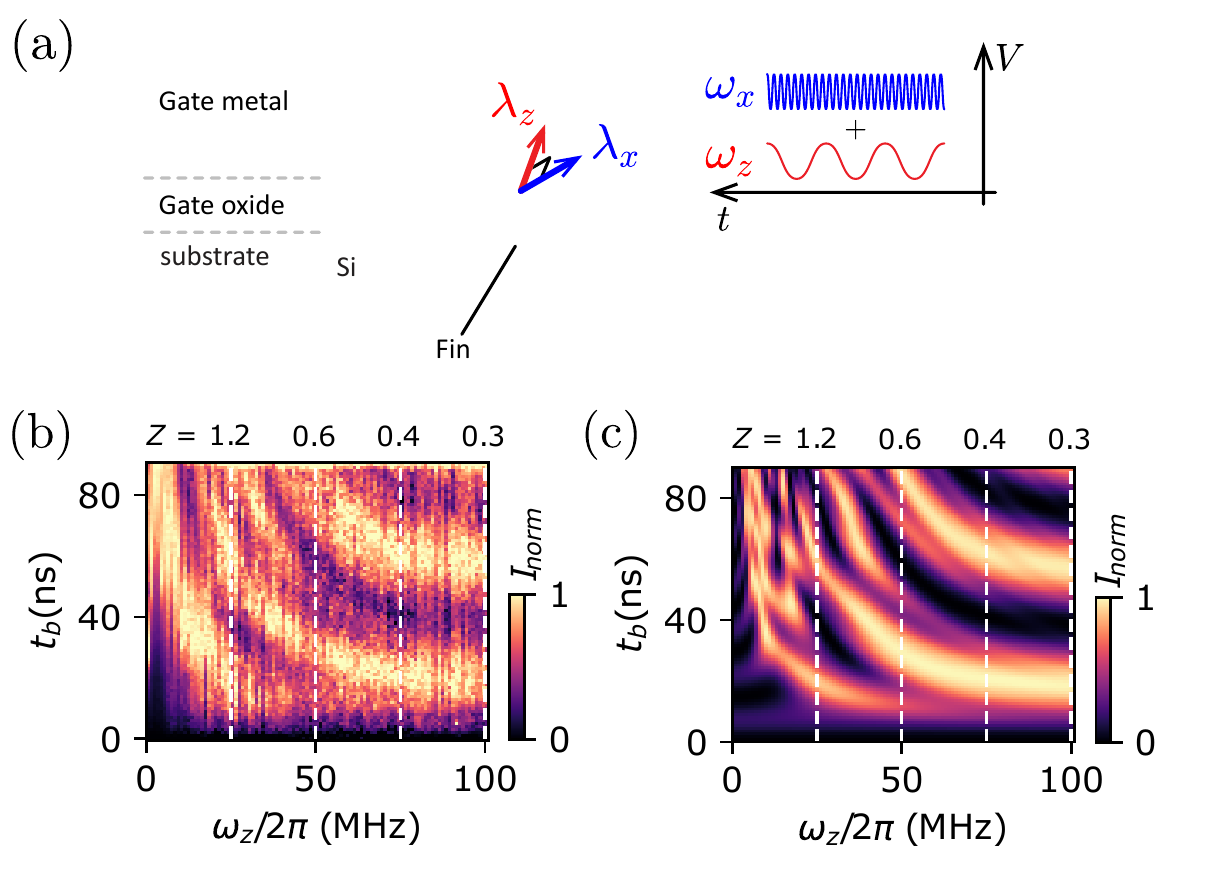}
\caption{(a) Schematic of our phase driven qubit. A hole spin qubit in a Si FinFET with Zeeman field $\textbf{b}$ is simultaneously driven by a microwave Rabi drive (blue) with amplitude $\lambda_x$ and frequency $\omega_x\approx\omega_q= |\textbf{b}|/\hbar$, and a radio-frequency phase drive (red) with amplitude $\lambda_z$ and frequency $\omega_z\ll \omega_q$. (b)-(c) Phase-driving-induced slow down of Rabi oscillations in qubit 1 (Q1). The Rabi frequency is suppressed as  $\omega_R=\lambda_x J_0(2Z)$, with $Z=\lambda_z/\omega_z$, and vanishes at $Z=1.2$. This prediction is experimentally confirmed in (b) by sweeping $\omega_z$ and measuring the Pauli-spin-blockaded current $I_\text{norm}$ normalized by the maximal current, proportional to the spin flip probability, for different burst times $t_b$. Regular oscillations are observed for $Z\lesssim 0.5$. For $Z\gtrsim 0.5$ phase driving causes non-trivial features that are captured by the simulation in (c) obtained from Eq.~\eqref{eq:H-full} with $\lambda_x/2\pi= 29$~MHz, $\lambda_z/2\pi= 30$~MHz and $\omega_q/2\pi = \omega_x/2\pi = 4.5$~GHz.}
\label{fig:slowdown}
\end{figure}

Single qubit operations rely on flipping spin states on demand.
A microwave pulse drives spin rotations, resulting in the well-known Rabi oscillations. 
For confined holes, these oscillations are fast and rely on either an electrically tunable and anisotropic $g$ tensor~\cite{doi:10.1063/1.4858959,PhysRevLett.120.137702,PhysRevB.98.155319} or  periodic spin motion in a SOI field~\cite{PhysRevB.74.165319,doi:10.1126/science.1148092,DRkloeffel1,DRkloeffel3}.
To date, however, these oscillations are qualitatively similar to competing qubit architectures~\cite{RevModPhys.93.025005,RevModPhys.75.281,Gilbert2023,Pla2012,Koppens2006}, and they occur at fixed microwave GHz frequencies determined by the qubit energy. Qubit responses to detuned frequencies are  associated to non-linearities in the coupling to the driving field~\cite{PhysRevLett.115.106802,PhysRevB.73.125304}.

In this work, we investigate the dynamics of a hole spin qubit hosted in a Si FinFET under simultaneous application of longitudinal (phase) and transverse (Rabi) drives at radio frequency $\omega_z$ and microwave frequency $\omega_x$, respectively, see Fig.~\ref{fig:slowdown}(a). We demonstrate that the rich microscopic physics of hole nanostructures  leads to a non-trivial response of the qubit state to these oscillating fields even at frequencies far detuned from the qubit energy. This anomalous response arises from a strong interplay between the phase and Rabi electrical drives in the \textit{linear} regime. More specifically, we show that by driving the qubit phase at radio frequencies $\omega_z$ in the MHz range, i.e., three orders of magnitude lower than the  microwave GHz-range Larmor frequency of the qubit $\omega_q$, we can controllably  (i) suppress the resonant Rabi oscillations at $\omega_x=\omega_q$ and (ii) revive them at sideband frequencies $\omega_x =\omega_q \pm m \omega_z$, with $ m$ $\in$ $\mathbb{N^+}$. We observe this behaviour reproducibly on two different devices. The suppression of Rabi frequency can be exploited to detect longitudinal interactions between spins and microwave resonators~\cite{PhysRevLett.129.066801}. In future qubit processors using a global high-frequency driving field,  sideband oscillations can  provide ways to selectively address individual qubits by radio-frequency MHz signals, relieving demanding technological challenges for designing large-scale high-frequency circuits.
We also predict that our two-tone drive protects Rabi oscillations from noise as the periodic phase driving  gaps  the Floquet spectrum of the system similar to the Bloch bands in periodic lattices. This noise suppression is a valuable tool for dressed spin qubit architectures~\cite{Laucht2017,PhysRevB.104.235411,PhysRevA.104.062415,doi:10.1063/5.0096467} and for Floquet engineering high-fidelity quantum gates~\cite{Xue2022,PhysRevA.100.012341,PhysRevA.98.043436,Yang2019,PhysRevA.99.052321,rimbach2022simple}.

\paragraph{Electrical manipulation of hole spins.|}
A hole spin qubit in an external magnetic field $\textbf{B}$ is described by the Zeeman Hamiltonian $H_q =\textbf{b}\cdot \pmb{\sigma}/2$, where $\textbf{b}=\mu_B \textbf{B}\cdot \hat{g}$ is the Zeeman vector,  $\hat{g}$ is the electrically-tunable $g$  tensor of the system, and $\pmb{\sigma}$ is the vector of Pauli matrices. Quite generally, an electrical pulse with frequency  $\omega$ applied to the system gives rise to an oscillating vector field $\pmb{\lambda}\cos(\omega t)$ that directly couples to the qubit via $H_d= \hbar\pmb{\lambda}\cdot \pmb{\sigma} \cos(\omega t)$ due to the SOI. The vector $\pmb{\lambda}\cos(\omega t)$ models the drive as a time-dependent Zeeman field acting on the qubit. Its direction depends on the microscopic details of the nanostructure, and includes processes such as $g$ tensor modulation and electric dipole spin resonance~\cite{PhysRevB.74.165319,DRkloeffel1,DRkloeffel3}, while its amplitude scales linearly with the applied microwave field. These processes enable electrical manipulation of qubits with multiple driving frequencies and amplitudes.  Transitions between spin up and down states occur for $\pmb{\lambda}\perp \textbf{b}$ (Rabi driving), while only the phase of the qubit is addressed~\cite{PhysRevLett.129.066801,PhysRevB.91.094517,PhysRevB.93.134501,PhysRevB.103.035301,PhysRevB.99.245306,Bottcher2022} for $\pmb{\lambda}\parallel \textbf{b}$ (phase driving). Interestingly,  while phase driving alone cannot induce  Rabi oscillations, a radio-frequency phase pulse can significantly alter the dynamics of the qubit when acting together with a Rabi driving field.

We stress that all of our findings lie within the scope of linear response in the driving field. Variations from this linear regime were detected at large driving powers~\cite{Yoneda2018,Froning2021,undseth2022nonlinear}. Non-linearities in the driving field $\propto \pmb{\lambda}^2\cos(\omega t)^2$ induced by excited states were also proposed as a source of higher-harmonic response in spin qubits~\cite{PhysRevLett.115.106802,PhysRevB.73.125304}. 

\paragraph{Phase driven qubits.|}
We consider a spin qubit with GHz-range frequency $\omega_q=|\textbf{b}|/\hbar$. A transverse Rabi drive $ \lambda_x \cos(\omega_xt)$ with amplitude $\lambda_x$ and frequency $\omega_x= \omega_q-\Delta$ induces Rabi oscillations when the MHz-range detuning $\Delta$ is small. This system exhibits Rabi oscillations  with frequency $\omega_R=\sqrt{\Delta^2+\lambda_x^2}$ and maximal spin-flip probability $P_R^\text{max}=\lambda_x^2/(\Delta^2+\lambda_x^2)$.
We add an additional simultaneous longitudinal phase drive $ \lambda_z \cos(\omega_zt)$, with amplitude $\lambda_z$ and frequency $\omega_z\sim$~MHz, that is far detuned from $\omega_q$. The two-tone  Hamiltonian reads
\begin{equation}
\label{eq:H-full}
H= \frac{\hbar\omega_q}{2}\sigma_z+ \hbar\lambda_x \sigma_x\cos(\omega_x t)+ \hbar\lambda_z \sigma_z\cos(\omega_z t) \ .
\end{equation}
The direction $\hat{z}$ ($\hat{x}$) is parallel (perpendicular) to $\textbf{b}$, Fig.~\ref{fig:slowdown}(a).

By moving to the rotating frame defined by the  transformation $ 
U_r(t)=e^{-i\sigma_z\left[\omega_x t+2Z \sin\left(\omega_z t\right)\right]/2 } $~\cite{PhysRevLett.129.066801}, which exactly accounts for the phase driving, and  neglecting terms rotating at frequencies $\sim 2\omega_x$, we obtain
\begin{widetext}
\begin{equation}
\label{eq:RWA-Fin}
\tilde{H}=\frac{\hbar\Delta}{2}\sigma_z+ \frac{\hbar\lambda_x }{2}J_0(2Z)  \sigma_x  
+ \hbar\lambda_x \sum_{n=1}^{\infty} \Big(J_{2n}(2Z)\cos[2n\omega_z t]\sigma_x - J_{2n-1}(2Z)\sin[(2n-1)\omega_z t]\sigma_y   \Big) \ ,
\end{equation}
\end{widetext}
with dimensionless parameter $Z=\lambda_z/\omega_z$ and  Bessel functions $J_n$.
Note that without phase drive, i.e., $\lambda_z=0\Rightarrow Z=0$, and since $J_0(0)=1$ and $J_{n\ne 0}(0)=0$, Eq.~(\ref{eq:RWA-Fin}) reduces to $\tilde H = \hbar(\Delta\sigma_z+  \lambda_x  \sigma_x)/2$, i.e.,  the rotating frame Hamiltonian for Rabi driven qubits in the rotating wave approximation (RWA). 

Close to resonance $\Delta\lesssim \omega_z,\lambda_x$, we obtain $J_0(2Z)=1-Z^2+\mathcal{O}(Z^4)$ and  the first correction to the qubit dynamics caused by the phase driving is $\propto Z^2$. Consequently, at moderate values of $\lambda_z$ and when $\omega_z\sim \omega_q$, then $Z\ll 1$, and the phase driving has no effect. Moreover,  Rabi pulses with $\omega_x\ll \omega_q$ are off-resonant and do not affect the qubit.
 These considerations justify using the Hamiltonian $H$ in Eq.~\eqref{eq:H-full} also in general cases where  $\lambda_{x}$ [$\lambda_z$] has an additional component parallel [perpendicular] to $\textbf{b}$.
Finally, a relative phase difference $\varphi$ between the two driving signals is relevant at comparable values of $\omega_x$ and $\omega_z$~\cite{PhysRevLett.129.066801}, but can be neglected when $\omega_z\ll\omega_x$. Finite $\varphi$'s become relevant in the presence of noise, as discussed later.

\paragraph{Sideband Rabi oscillations.|} 

\begin{figure*}[t]
\centering
\includegraphics[width=1.5\columnwidth]{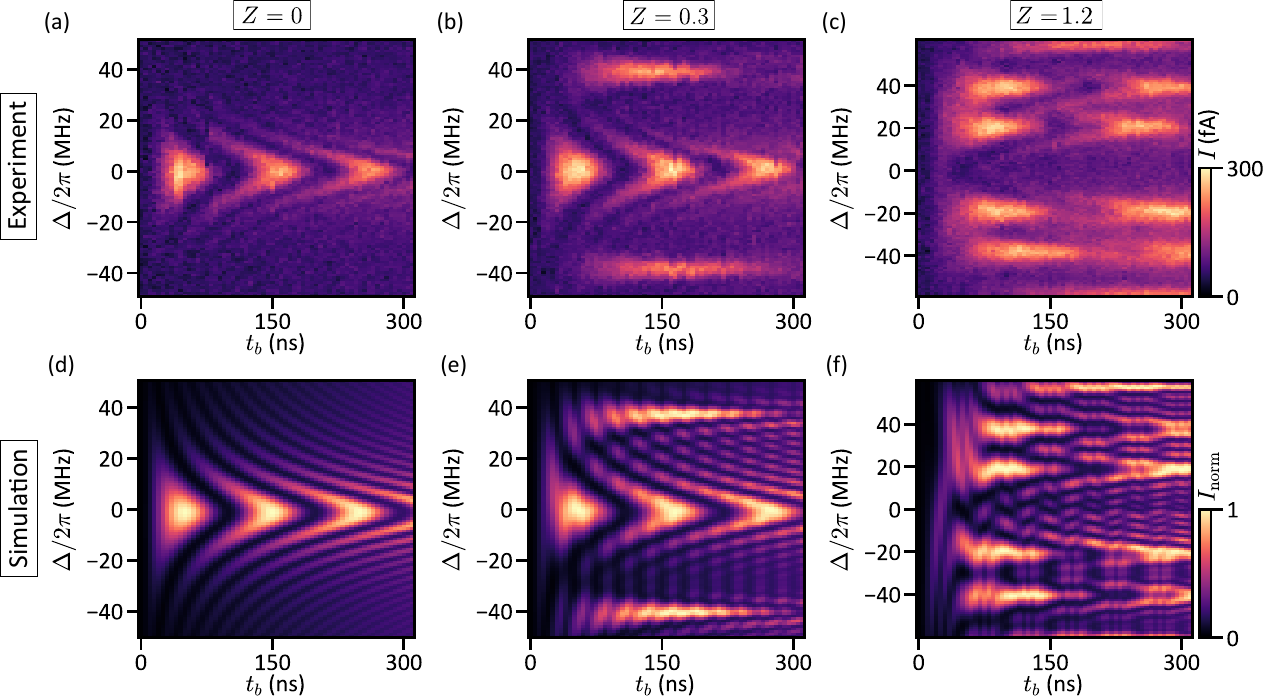}
\caption{Sideband Rabi oscillations in qubit 2 (Q2). (a),(b),(c) Spin precession for simultaneous microwave Rabi driving and radio-frequency  phase driving against burst time $t_b$ and detuning $\Delta$. We show results for $Z=0$, $Z=0.3$, and $Z=1.2$. The typical chevron pattern centred at $\Delta=0$ in (a) is modified in (b),(c) by additional sidebands at $\Delta/2\pi = \pm n\cdot\omega_z/2\pi=\pm n\cdot40$\,MHz  and $\pm n\cdot 20$\,MHz, respectively. The $\Delta=0$ oscillations are slower in (b) and completely vanish in (c). (d),(e),(f) Simulations of the time-evolution governed by Eq.~\eqref{eq:H-full}, showing excellent agreement with experiments. We use $\omega_q/2\pi=4.95\,$GHz, $\lambda_x/2\pi=10$\,MHz for the three $Z$ values and $\lambda_z=0$, $\lambda_z/2\pi=0.3 \omega_z/2\pi= 12$\,MHz, $\lambda_z/2\pi=1.2\omega_z/2\pi=24$\,MHz for (d),(e),(f), respectively. }
\label{fig:SBchevron}
\end{figure*}

We study the influence of the phase drive on the  oscillations for resonant ($\Delta=\omega_q-\omega_x=0$) and non-resonant ($\Delta \neq 0$) Rabi drives.
In the resonant case and when $\omega_z \gtrsim \lambda_x$, one can simplify $\tilde{H}$ in 
Eq.~\eqref{eq:RWA-Fin} by the RWA. 
The dominant contribution to $\tilde{H}$ is the static $n=0$ component, yielding fully developed oscillations with frequency $\omega_R=\lambda_x J_0(2Z)$,   Fig.~\ref{fig:slowdown}(b),~(c).

We verify this prediction experimentally in a hole spin qubit in two different Si FinFETs described in detail in Refs.~\cite{doi:10.1063/5.0036520,camenzind2021spin,geyer2022two}. 
Our first (second) qubit Q1 (Q2) is operated at  $\omega_q/2\pi = 4.5$\,GHz ($\omega_q/2\pi = 4.95$\,GHz) corresponding to a $g$ factor $g=2.14$ for $B=0.15$\,T ($g=2.72$ for $B=0.13$~T); $g$ depends on the gate potential $V$  with sensitivity $\partial g/\partial V\approx -0.05$\,V$^{-1}$ ($\partial g/\partial V\approx 0.41$\,V$^{-1}$). Our qubits  are initialized and read out via Pauli spin blockade and direct current integration, see Refs.~\cite{camenzind2021spin,geyer2022two}.
Our system enables high-bandwidth phase driving via the electrically tunable $g$ tensor and Rabi driving via electric-dipole spin resonance. These contributions are generated by applying two oscillating electrical signals at different frequencies,  Fig.~\ref{fig:slowdown}(a). Generally these tones induce both Rabi and phase driving, however, as discussed before, we discard the negligible contributions of far-detuned Rabi driving and nearly-resonant phase driving.

For the measurements in Fig.~\ref{fig:slowdown}(b), we apply simultaneously a resonant Rabi drive with $\omega_x=\omega_q$ and a phase drive with variable, far-detuned frequency $\omega_z$ to Q1. We measure the qubit state after the burst time $t_b$. The two pulses have comparable amplitudes, $\lambda_x/2\pi\approx\lambda_z/2\pi\approx 30$\,MHz.  Rabi oscillations can be observed along the vertical axis of the figure, and by sweeping $\omega_z$, we map out the dependence of $\omega_R$ on $Z=\lambda_z/\omega_z$. We find good agreement between our measurement and the quantum dynamics simulated by using  $H$ in Eq.~\eqref{eq:H-full}. We consistently reproduce this behaviour in Q2 [see the Supplemental Material~\cite{SM}].

Remarkably, Rabi oscillations are suppressed at certain values of $Z=Z_j$, defined by the roots of the Bessel function $J_0(2Z_j)=0$, where $\omega_R$ vanishes. The first root  $Z_1\approx 1.2$  corresponds to $\lambda_z\approx 1.2 \omega_z$ and can be observed in our experiment. At $Z=1.2$, the higher harmonic components in $\tilde{H}$ in Eq.~\eqref{eq:RWA-Fin} with $n\geq 1$ dominate the dynamics.

These higher harmonics are crucial to understand the Rabi sidebands appearing in the non-resonant case, comprising a finite detuning $\Delta$, shown in Fig.~\ref{fig:SBchevron}(b),(c). At small values of $\Delta\ll \lambda_x$, the Rabi frequency increases as $\omega_R=\sqrt{\Delta^2+\lambda_x^2J_0(2Z)^2}$, resulting in the typical chevron pattern, and suppressing the oscillations at large $\Delta$. However,  when $\lambda_z\approx \omega_z$, oscillations are revived at finite values of $\Delta$. In particular, at $\Delta= \pm m \omega_z$, the system is resonant with the $m^\text{th}$-harmonic in  Eq.~\eqref{eq:RWA-Fin} ($m\in \mathbb{N^+}$), and sideband  oscillations  at frequencies $\omega_R^m=\lambda_x J_m(2Z)$ are restored. 
 
In Fig.~\ref{fig:SBchevron}, we show measurements and simulations of Rabi oscillations against $\Delta$ at different values of $Z$ for Q2. 
The  Rabi chevron in (a) is modified by phase driving.
In (b), we consider $Z=0.3$, and we observe the appearance of sideband resonances at frequencies $\pm \omega_z/2\pi=\pm 40$\,MHz. In (c), by reducing $\omega_z$ to $\omega_z/2\pi=20$\,MHz and increasing $\lambda_z$, we reach $Z=1.2$, where the resonant  oscillations at $\Delta=0$ vanish and  only sideband oscillations remain. As shown in (d)-(f), these sidebands are well explained by our model, which is linear in the driving field amplitudes. 
We emphasize  that in contrast to non-linear driving, where the $\Delta=0$ resonance does not disappear and sidebands oscillations appear at fixed frequencies $\omega_x= \omega_q/m$ with $m\in \mathbb{N}^+$, by operating our qubit at $Z=1.2$ we completely remove the $\Delta=0$ oscillations and still fully control the sideband frequencies $\Delta=\pm m\omega_z=m \lambda_z/1.2$ by varying the  amplitude $\lambda_z$ of the radio-frequency pulse.

Our driving scheme opens the possibility of dynamically shifting the qubit frequency to higher harmonics, thus reducing frequency-crowding issues in dense large-scale quantum processors and enabling individual qubit addressability in global microwave fields by technologically inexpensive MHz circuits.

\paragraph{Effects of noise.|}

\begin{figure}[t]
\centering
\includegraphics[width=\columnwidth]{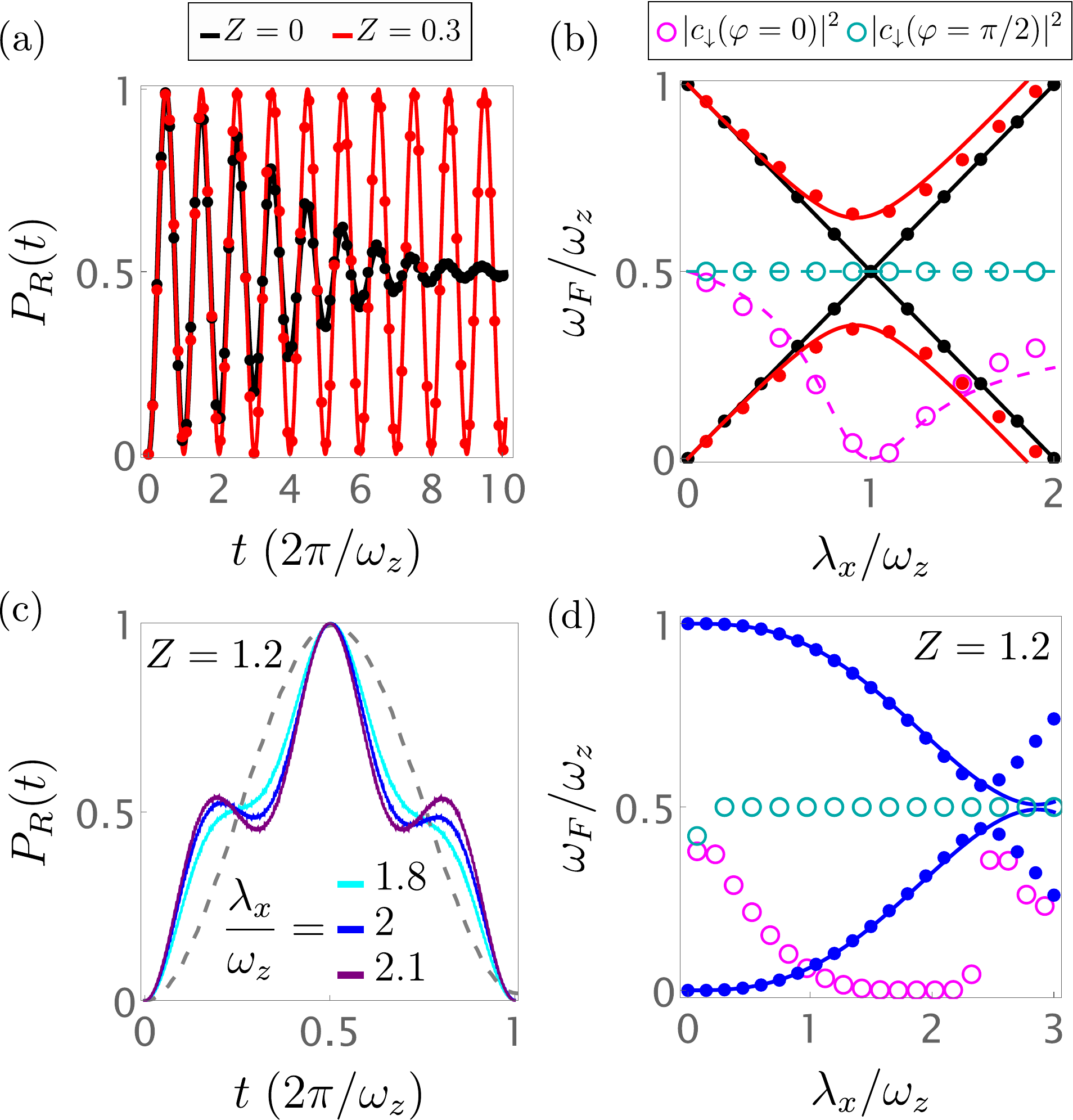}
\caption{(a) Simulated phase-driving-induced undamped Rabi oscillations at $\omega_z=\lambda_x$. Black (red) dots denote Rabi probability averaged over $N=10^3$ realizations of the noise Hamiltonian $H_N$ for $\lambda_z=0$ ($\lambda_z=0.3\omega_z$) picked from a Gaussian distribution with zero mean and standard deviation $\bar{\sigma}=0.1\omega_z$. The results match Eq.~\eqref{eq:Noisy-rabi} with $T_2^R=2\sqrt{2}\hbar/\bar{\sigma}$ ($T_2^R\to\infty$). (c)~Noisy Rabi oscillation at $Z=1.2$ and various  $\lambda_x/\omega_z$ ratios. The gray dashed line corresponds to the red line in~(a).
(b),(d) Floquet spectra against $\lambda_x/\omega_z$. (b) Solid black (red) dots show numerically computed Floquet energies $\omega_F$  at $\lambda_z=0$ ($\lambda_z=0.3\omega_z$); the lines denote the approximation in Eq.~\eqref{eq:Floq.en}. Magenta (cyan) circles exhibit the probability $|c_\downarrow|^2=|\langle\downarrow|0_F\rangle|^2$ for a two-tone relative phase $\varphi=0$ ($\varphi=\pi/2$). Dashed lines show Eq.~\eqref{eq:Floq.c}. (d) Blue dots show the Floquet spectrum at $Z=1.2$, while the lines display the fitting formula $\omega_F^{0,1}=\pm\omega_z[0.084 (\lambda_x/\omega_z)^3-0.022(\lambda_x/\omega_z)^4]~\text{mod}(\omega_z)$.
We use $\omega_q=10^3\omega_z$.
}
\label{fig:Floquet}
\end{figure}

The coupling to the environment causes decoherence and damps the Rabi oscillations in a time $T_2^R$, see Fig.~ \ref{fig:Floquet}(a), black curve. The Rabi decay time $T_2^R$ is dominated by noise at frequencies close to the driving frequency $\omega_x$. We model this phenomenologically by the noise Hamiltonian $H_N= \cos(\omega_x t) \textbf{h}\cdot \pmb{\sigma}/2 $, describing the spin coupling to a stochastic Gaussian distributed vector $\textbf{h}$, with zero mean and diagonal covariance matrix $\Sigma_{ij}=\delta_{ij}\bar{\sigma}^2 $. This model accurately describes high-frequency noise form different sources~\cite{Yang2019,PhysRevLett.127.190501,PhysRevLett.116.066806,Wang2021}.

After ensemble averaging and focusing on the resonant case $\Delta=0$, we find that for conventional Rabi driving, $H_N$ suppresses the Rabi oscillations as
\begin{equation}
\label{eq:Noisy-rabi}
P_R(t)=\frac{1}{2}\left[1-e^{-(t/T_2^R)^2}\cos(\omega_R t)\right] \ , 
\end{equation} 
with $T_2^R=2\sqrt{2}\hbar/\bar{\sigma}$.
The decay time $T_2^R$ is larger than the dephasing time $T_2^*$ and determines the lifetime of dressed spin qubits~\cite{Laucht2017,PhysRevB.104.235411,PhysRevA.104.062415,doi:10.1063/5.0096467}, utilizing nearly-resonant always-on global microwave fields. However, when the SOI is large, $T_2^R$ is significantly shortened, and becomes comparable to $T_2^*$~\cite{Wang2022,Froning2021}, thus limiting the advantages of these architectures.

\paragraph{Protection of Rabi oscillations from noise.|}
We simulate the effect of an additional phase pulse with $\lambda_z\ll \lambda_x$ at frequency $\omega_z \approx \lambda_x$.  
As shown in Fig.~\ref{fig:Floquet}(a), even a small phase driving (red curve) decouples Rabi oscillations from noise and enhances  $T_2^R$ by orders of magnitude. We also verified that these decay-free oscillations are robust against noise at different frequencies.

The origin of persistent oscillations can be understood in terms of the Floquet modes $|0_F\rangle, |1_F\rangle$~\cite{doi:10.1080/00018732.2015.1055918,Rudner2020}. These are eigenstates of the Floquet operator  $U(T)=U_r(T)\mathcal{T}e\left[-i\int_0^{T}\tilde{H}(\tau)d\tau/\hbar \right]$ with eigenvalues $e^{-i\omega_F^{0}T}, e^{-i\omega_F^{1}T}$, respectively. Here, $\mathcal{T}e$ is the time-ordered exponential, the period $T=2\pi/\omega_z$, and  $U_r$ transforms the system back to the lab frame~\footnote{To use Floquet theory, we assume that $\omega_x=\omega_q$ is commensurate with $\omega_z$. However our results are valid generally when $\omega_x\gg \omega_z,\lambda_x,\lambda_z$, $\omega_x$ and fast oscillating corrections $\propto 1/\omega_x$  are negligible.}. 

The eigenvalues and eigenvectors of $U(T)$ are shown in Fig.~\ref{fig:Floquet}(b).
First, in contrast to the usual Rabi driving, when a phase driving pulse is applied at frequencies comparable to $\lambda_x$ and is in-phase with the Rabi drive, the spin states  in the lab frame coincide with the Floquet modes, i.e. $|\uparrow\rangle=|0_F\rangle$, $|~\downarrow~\rangle=|1_F\rangle$ (magenta line). 
Second, phase driving opens a gap of size $\omega_F^1-\omega_F^0\approx\lambda_z$ in the Floquet spectrum (black and red lines), that protects the system from moderate noise sources with $\bar{\sigma}\lesssim \lambda_z$. For  $\lambda_z\ll \lambda_x \sim \omega_z$, the Floquet eigenenergies are 
\begin{equation}
\label{eq:Floq.en}
\omega_F^{0,1}=\pm\frac{1}{2}\left[\omega_z+\sqrt{(\lambda_x-\omega_z)^2+\frac{\lambda_x^2}{\omega_z^2}\lambda_z^2}\right] \text{mod}(\omega_z)\ .
\end{equation}

In analogy to disorder potentials in Bloch bands, when the system is initialized in an eigenstate, transitions to other eigenstates are suppressed as long as the standard deviation of the disorder is smaller than the energy gap. 
This comparison allows us to identify the decay-free  oscillations shown in red in Fig.~\ref{fig:Floquet}(a) as the temporal evolution of an individual Floquet mode. 

The decay-free Rabi oscillations depend on the gapped Floquet spectrum and  the possibility of preparing a Floquet eigenmode. 
In Fig.~\ref{fig:Floquet}(b), we show  that a relative, experimentally-tunable phase $\varphi$ between the Rabi and phase tones can be used to select arbitrary superpositions of Floquet states (magenta and cyan curves).
The amplitudes $c_s=\langle s|0_F\rangle$ between the spin state $|s=\uparrow\downarrow\rangle$ and the Floquet state $|0_F\rangle$ are
\begin{equation}
\label{eq:Floq.c}
c_{\uparrow\downarrow}(\varphi)\approx \frac{\cos(\theta)\pm \sin(\theta) e^{i\varphi} }{\sqrt{2}} \ , \ \tan(2\theta)=\frac{\lambda_x\lambda_z}{\omega_z|\lambda_x-\omega_z|} \ .
\end{equation}

For strong phase drivings, where $\lambda_z\sim \lambda_x$, Eqs.~\eqref{eq:Floq.en} and~\eqref{eq:Floq.c} are inaccurate, but simulations still predict decay-free oscillations and gapped Floquet spectrum. 
For example,  in Fig.~\ref{fig:Floquet}(c),(d) we examine the oscillations and the Floquet spectrum at $Z=1.2$. 
The Floquet bands touch at $\lambda_x=0$ $\text{mod}(\omega_z)$; at $\lambda_x\gtrsim \omega_z$, the gap becomes significant and decay-free oscillations persist for a wide range of parameters.
Because of the strong phase driving the oscillations are not sinusoidal. Their peculiar shape probes the temporal structure of the Floquet mode and is shown in Fig.~\ref{fig:Floquet}(d) for different values of $\lambda_x$.

We note that relaxation between Floquet modes is suppressed, as reflected in persistent oscillations, but superpositions of Floquet modes are still subjected to dephasing with characteristic time $T_2^R$. 
We envision that the possibility to stabilize Floquet modes by phase driving  opens a wide range of exciting opportunities to optimize dressed qubits, and to prepare exotic states in future Floquet metamaterials.

\paragraph{Conclusion.|}
We demonstrated  that radio-frequency phase driving of hole spin qubits induces collapse and revival of Rabi oscillations, resulting in oscillations at sidebands of the qubit frequency. These sidebands do not require non-linear coupling of the spin to the driving field. We show theoretically that phase driving  also leads to decay-free Rabi oscillations in noisy qubits.
Our two-tone driving scheme provides an alternative way of implementing individual addressability in global microwave fields in future large-scale qubit architectures, and  Floquet engineering high-fidelity qubit gates.

\paragraph{Acknowledgements.|}
We thank T. Berger for the rendering of our FinFET device.
We acknowledge support by the cleanroom operation team, particularly U. Drechsler, A. Olziersky, and D. D. Pineda, at the IBM Binnig and Rohrer Nanotechnology Center, and technical support at the University of Basel by S. Martin and M. Steinacher. 
L.C.C. acknowledges support by a Swiss NSF mobility fellowship (P2BSP2200127).
This work was supported by the Swiss National Science Foundation, NCCR SPIN (grant number 51NF40-180604) and  by the EU H2020 European Microkelvin Platform (EMP) Grant No. 824109.

\bibliography{references}

%%%%%%%%%%%%%%%% SUPPLEMENTARY
\clearpage
\newpage
\mbox{~}
%\clearpage
%\newpage

\onecolumngrid

\begin{center}
  \textbf{\large Supplemental Material:\\ Phase driving  hole spin qubits}\\[.2cm]
\end{center}

\setcounter{equation}{0}
\setcounter{figure}{0}
\setcounter{table}{0}
\setcounter{section}{0}

\renewcommand{\theequation}{S\arabic{equation}}
\renewcommand{\thefigure}{S\arabic{figure}}
\renewcommand{\thesection}{S\arabic{section}}
\renewcommand{\bibnumfmt}[1]{[S#1]}
\renewcommand{\citenumfont}[1]{S#1}

\maketitle

\section{Additional measurements of our first and second devices}

\begin{figure}[h]
\centering
\includegraphics[width=0.7\columnwidth]{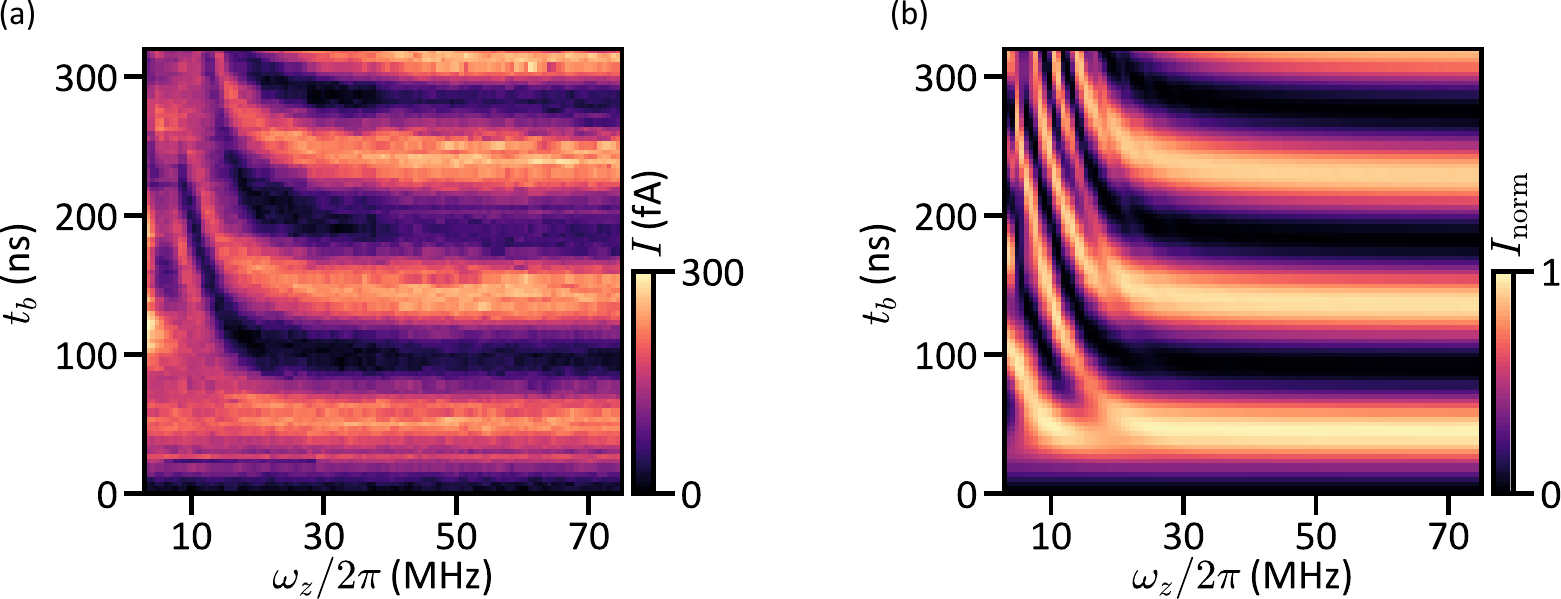}
\caption{
Phase-driving-induced slowing down of Rabi oscillations in qubit 2 (Q2). Measurements (a) and simulations (b) match well, and this trend is analogous to the one reported in the main text for Q1, see Fig.~1(b),(c).
Here, we used $\omega_x/2\pi = \omega_q/2\pi = 3.115$\,GHz,  $\lambda_x/2\pi=11$~MHz, and $\lambda_z/2\pi=6.1$~MHz.
}
\label{fig:slowdown}
\end{figure}

\begin{figure}[h]
\centering
\includegraphics[width=0.7\columnwidth]{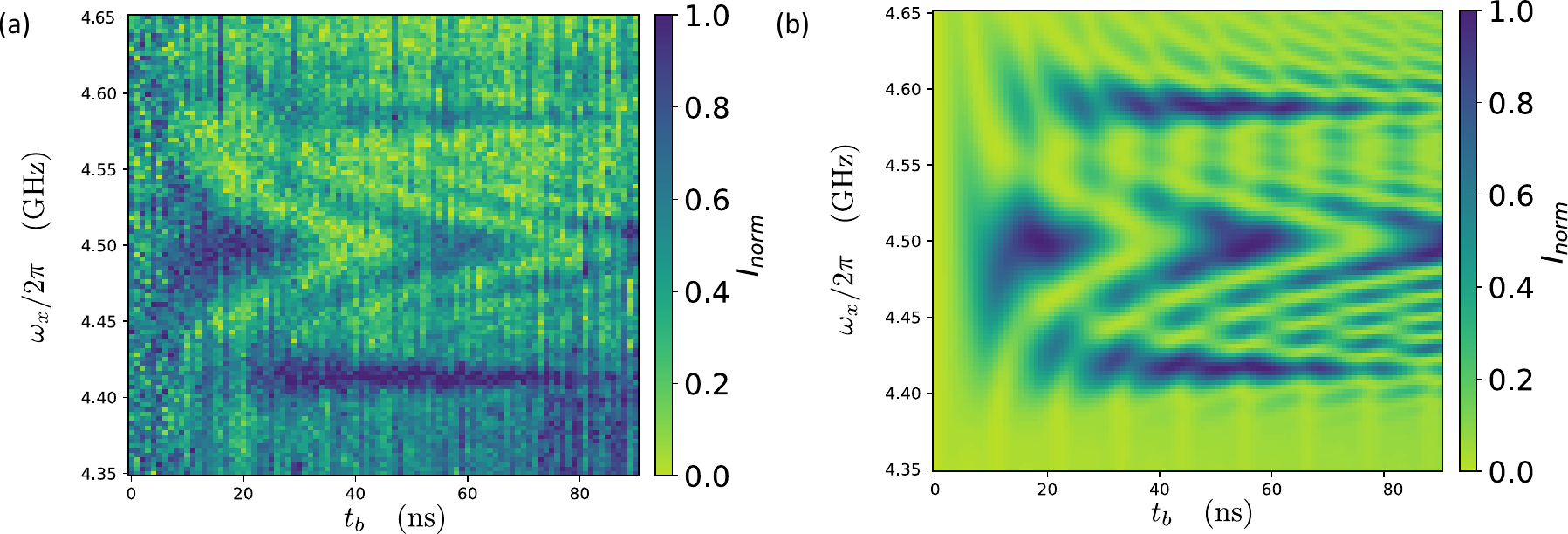}
\caption{
Phase-driving-induced sideband Rabi oscillations in qubit 1 (Q1). Measurements (a) and simulations (b) are in good agreement, and these sidebands are analogous to the ones reported in the main text for Q2, see Fig.~2(b),(e).
Here, we used $\omega_q/2\pi = 4.5$\,GHz,  $\lambda_x/2\pi=\lambda_z/2\pi=30$~MHz, and $\omega_z/2\pi=90.5$~MHz, corresponding to $Z=0.33$.
}
\label{fig:side}
\end{figure}

We present here additional data from our two qubits, Q1 and Q2, encoded in two different devices.
In Fig.~\ref{fig:slowdown}, we show the slowing down of Rabi oscillations by phase driving Q2. Compared to  Figs.~1(b),(c) in the main text, we observe a similar trend, with a lower Rabi and phase driving amplitudes. The measurement in  Fig.~\ref{fig:slowdown}(a) matches well the numerical simulation in Fig.~\ref{fig:slowdown}(b).
In Fig.~\ref{fig:side}, we show phase-driving-induced sideband oscillations appearing at finite detuning in Q1. These results are comparable to the ones obtained for Q2 and shown in Figs.~2(b),(e) in the main text. Also in this case, we observe a good agreement between measurements (a) and simulations (b).

\end{document}